\documentclass[floatfix,preprint,onecolumn,showpacs,aps,pre]{revtex4-1}
\usepackage{hyperref}
\usepackage[T1]{fontenc}
\usepackage{amssymb}
\usepackage{amsmath}
\usepackage{graphicx}
\usepackage{subfigure}
\usepackage{etoolbox}
\apptocmd{\thebibliography}{\raggedright}{}{}

\begin{document}
\title{Spreading processes with a unique absorbing state and finite lifetime of populations}
\date{\today}
\author{Norbert Barankai}
\author{J\'{o}zsef St\'{e}ger}
\affiliation{Department of the Physics of Complex Systems, E\" otv\" os
University, P\'azm\'any P\'eter S\'et\'any 1/A, H-1117 Budapest, Hungary}
\begin{abstract}
The presence of a unique absorbing state in the finite state space of an epidemic process always poses a challenge in the definition of its epidemiological threshold. Without the elimination of the absorbing state of the SIS process we show that an exponentially distributed finite lifetime of the population results in a threshold like beahivour in the long time limit. We illustrate our findings with the SIS dynamics on the complete graph and the star.
\end{abstract}
\maketitle

\section{Introduction}
\label{SEC1}
Stochastic modeling is a common way to simulate real world spreading processes like biological epidemic outbreaks, information spreading in social media, virus spreading in computer networks, etc.~(see \cite{EpidemicReview} and references therein). In the case of epidemiology the current state of the art of the field enables researchers to supply decision makers of national and international agencies with predictions originated in large scale computer simulations \cite{Influenza}. Building an epidemic simulation \cite{KeelingRohani} often consists of a choice of an epidemic model describing the various internal states of an individual and a prescription of dynamics that describes the state changing rules of individuals. An appropriate choice of parameters of the dynamics is also inevitable.

Consider a SIS process \cite{EpidemicReview,MoezMassouile} taking place in a population of humans. The internal states of the individuals can be susceptible (S) and infected (I). The heterogenous structure of the social life of the population  is modeled by a simple, unweighted graph $\mathcal{G}$. A state of the population $\omega$ is an assignment of internal states to the individuals. If it is fixed the following competing processes can take place. Either an infected individual halts and becomes susceptible or a susceptible individual also becomes infected if it is in connection with at least one infectious agent. The rates of these processes are called $\delta$ curing rate and $\beta$ infection rate. It is clear that different choices of $\delta$ and $\beta$ on the same graph must lead to markedly different dynamics. In order to implement the time evolution it is natural to choose a continuous time Markov process whose state space is the collection of the possible epidemic states of the population. The rate of a transition $\omega\rightarrow\omega'$ is nonzero if and only if the two states differ in the internal state of only a single individual such that this individual is cured or suffers infection from an infected neighbor. The transition rate is $\delta$ in the first case and the number of infectious neighbors multiplied by $\beta$ in the second one. 

With a mathematical model being at hand, one can ask for the the long time asymptoic state of the system. The following scenario of the parameter dependence of the model is usually demanded. By fixing the curing rate $\delta$ and the graph $\mathcal{G}$ the long time limit of the number of infected individuals has to be zero for small values of $\beta$. After increasing $\beta$ one achieves a threshold $\beta_c(\delta,\mathcal{G})$ beyond which there must be a finite ratio of infected individuals in the long time limit. Above the threshold, the density of the infected individuals must be a strictly monotone function of $\beta$. Although the closed form of the time evolution is generally not accessible due to the complexity of the problem, several mean field methods \cite{PastorSatorrasVespignani2001,EpidemicReview,VanMieghemOmicKooij2009} have been applied to determine $\beta_c(\delta,\mathcal{G})$ when the population has only a finite number of members. It turned out that the epidemic threshold satisfies the inequality \cite{Wangetal,VanMieghemOmicKooij2009,CatorVanMieghem2012,VanMieghem2012}
\begin{equation}
\frac{\delta}{\lambda_1(A[\mathcal{G}])}\leq \beta_{c}(\delta,\mathcal{G}),
\label{INEQ}
\end{equation}
where $\lambda_1(A[\mathcal{G}])$ is the largest eigenvalue of the adjacency matrix of $\mathcal{G}$ \cite{ChungBook,VanMieghemBook}. The inequality becomes strict in the case of complete graphs resulting in $\beta_{c}(\delta,K_N)=\delta/(N-1)$. For general graphs, there is a sequence of lower bounds improving \ref{INEQ}, see \cite{CatorVanMieghem2012} for details.

Unfortunately when one aims at performing a similar analyzis within the theoretical framework of Markov processes one immadiatelly runs into a serious problem if the population is finite \cite{VanMieghemCator2012,CatorVanMieghem2013}. To see this, observe that the SIS process scetched below has a single absorbing state. It is the state that contains only susceptible individuals. Such a unique absorbing state has a profound consequence on the asymptotic probability distribution of the Markov process. Namely, for sufficiently large times, the probability of finding the system in the absorbing state is almost one making \cite{VanMieghemOmicKooij2009} it impossible to define $\beta_c(\delta,\mathcal{G})$ through the calculation of the expectation number of infected individuals by the asymptotic state of the Markov process.

This phenomenon is a direct consequence of the finiteness of the phase space of the Markov process and resonates to the general experience of condensed matter physics, which tells that there are no phase transition or any other non-analytic behaviour in finite systems. This shows that if one would like to observe non-analytic behaviour in a SIS process without modifying the dynamics one has to perform a limit in which the size of the population tends to infinity.

In this paper we will follow another approach which we found when we started to study epidemic processes in a population where groups can form and decay i.e.~the connections between individuals have finite lifetime. The study of such processes \cite{Stehleetal2011,Machensetal2013} is becoming more and more important as mesoscale interaction data among humans are becoming avaliable to researchers (see \cite{Zhaoetal2011} and references therein). If groups of a population have finite lifetime then the definition of the epidemic process needs a modification, namely, we have to introduce stopping time which is also a random variable with a given probability distribution. What we have found is that (at least in the case of the complete graph and the star which we studied closely) the $\beta$ dependence of the expectation value of the final number of the infected individuals in a SIS process mimics the behaviour mentioned in the previous paragraph. It is clear that such an investigation is important in order to understand epidemic processes in a population capable of showing random group formation phenomena. 

Our viewpoint is very close to the intuitive picture presented in \cite{CatorVanMieghem2013} where the authors eliminate the absorbing state of the SIS process and analyze the resulting modified dynamics as the steady state with decay rate $\delta$ of the original process. Unfortunately, they do not detail how this picture can be formalized without the appearance of the trivial stationary distribution. Our work can be thought as a further step in this direction.

This paper is organized as follows. In Section \ref{SEC2} we introduce the epidemic processes under consideration. In Section \ref{SEC3} we show that in order to calculate the average final probability distribution of an epidemic process taking place in a population with an exponentially distributed finite lifetime we have to calculate the resolvent of the infinitesimal generator of the process. We use this result in Section \ref{DISCUSSION} to have a closer look on the SIS process taking place in the complete graph and the star. Finally we present some concluding remarks and outlooks regarding our work in  Section \ref{CONCLUSION}. 

\section{Epidemic processes on the complete graph and the star}
\label{SEC2}
\begin{figure}[!ht]
\centering
\includegraphics[scale=2.3]{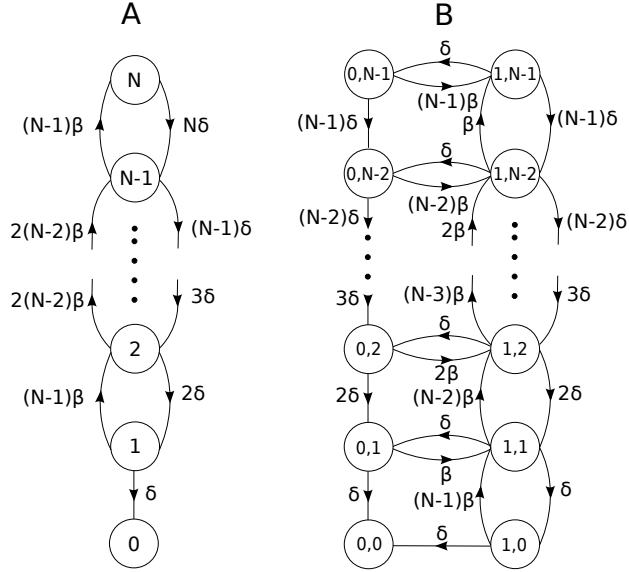}
\caption{State space and transition rates of the lumped SIS process on the complete graph $K_N$ (A) and the star $S_N$ (B)}
\label{FIG1}
\end{figure}

Thanks to the high symmetry of the complete graph $K_N$ and the star $S_N$ (which in this paper consists of a center node and $N-1$ leaves), the continous time Markov process of a SIS dynamics is lumpable \cite{Simonetal2011} i.e.~the states of the process can be joined together into mutually disjoint sets such that if these sets are thought to be states of a stochastic process, it will be Markovian too. In the case of $K_N$ the lumping process results in a state space $\mathcal{S}=\{0,1,\dots,N\}$, where a given state is nothing else but the total number of infected individuals of the system. The lumped state space of the SIS dynamics taking place on $S_N$ is $\mathcal{S}=\{(s,0),(s,1),\dots,(s,N-1):s=0,1\}$, where $s$ is the state of the central node and the second member of each tuple describes the number of infected individuals on the perimeter of the star such that in the state $(s,k)$, $s+k$ gives the total number of infected individuals. The state space and the transition rates of the SIS process on $K_N$ and $S_N$ can be seen in Fig.~\ref{FIG1}. It is clear that both  processes have a unique absorbing state charachterized by the absence of infected individuals. Removal of the unique absorbing state leads to the modified SIS process (mSIS) \cite{CatorVanMieghem2013}. Note that since the mSIS dynamics has no absorbing state its asymptotic distribution has a nonuniform dependence on $\beta,\delta$ and $\mathcal{G}$ \cite{CatorVanMieghem2013}.

\section{Effect of finite lifetime on spreading processes}
\label{SEC3}
\subsection{Illustrative example: the SI process on the complete graph}
\begin{figure}[!ht]
\centering
\includegraphics[scale=0.65]{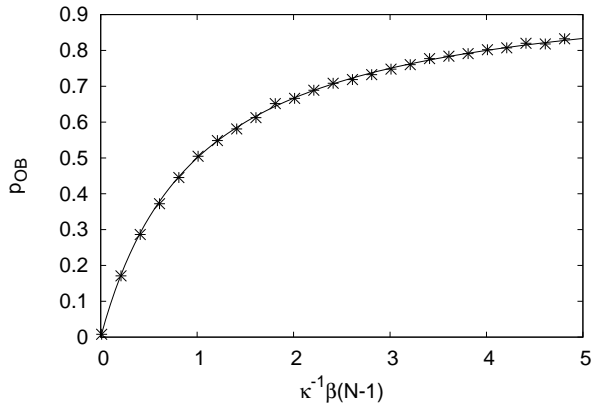}
\caption{Probability of an outbreak in SI processes driven on $K_{50}$ initiated by a single infected individual. Theoretical curve of Eq.~\ref{EQ:SI} as well as results of simulations are presented.}
\label{FIG2}
\end{figure}

The SI process is the $\delta\rightarrow 0$ limit of the SIS process. In a homogeneous group of $N$ individuals the lumped time evolution is a simple Markov chain with one isolated and one absorbing state. The absence of infected individuals is the isolated state whereas the absence of susceptible agents is the absorbing one. Consider $0<n<N$, the initial number of infected individuals in the population. For a duration much longer than $(\beta n(N-n))^{-1}$ the system will be in the absorbing state with high probability. That means the number of infected individuals will be $N$ with probability $1$ when $t\rightarrow \infty$ independently of the value of $\beta$. On the other hand when the group has finite lifetime $\tau$ the situation changes: in finite time the finite Markov chain (thought as a hold and jump process with exponentially distributed holding times)  cannot reach the absorbing state with probability one. Instead, the probability that at least one infection takes place before the group dissolves is $1-\exp{(-\beta n(N-n)\tau)}$, which we identify with the probability of an outbreak happening during $\tau$. When the lifetime of the group has an exponential probability distribution $\mathrm{Exp}(\kappa)$, the probability of an outbreak in a population containing initially $n$ infected individuals becomes a random variable with an expectation value
\begin{align}
p_{\mathrm{OB}}(n)&=\int_{0}^{\infty}(1-e^{-\beta n(N-n) \kappa})\kappa e^{-\kappa\tau}d\tau\nonumber\\
&=\frac{\beta n(N-n)}{\beta n(N-n)+\kappa},\label{EQ:SI}
\end{align}
which is definitely less than $1$ unless $\kappa$ vanishes i.e.~$\tau\rightarrow\infty $ (see Fig.~\ref{FIG2}). 

\subsection{Enumerable Markov processes with exponentially distributed running time}
Consider a continuous time Markov process with a countable state space $\mathcal{S}$. The process is completely determined by the transition rates $0\leq q_{ij}$ i.e.~the rate of an $i\rightarrow j$ transition between nonequal states $i$ and $j$ of $\mathcal{S}$. The infinitesimal generator $Q$ of the process is defined by
\begin{equation*}
Q_{ij}=
\begin{cases}
q_{ij} & \text{if } i\neq j\\
-\sum_{j\neq i} q_{ij} & \text{if } i=j\\
\end{cases}
\end{equation*}
and is assumed to be bounded \footnote{The boundedness of the generator is always guaranteed when $\mathcal{S}$ is finite or the magnitude of entries of $Q$ and the number of nonzero entries in each row are uniformly bounded.}. If the initial probability distribution is represented by the vector $\mathbf{p}(0)$ then the time evolution is governed by the exponential of $Q$~\cite{Norris}:
\begin{equation*}
\mathbf{p}^T(t)=\mathbf{p}^T(0)e^{Qt}.
\end{equation*}
Assume that the running time $\tau$ of the process is a random variable with an exponential distribution $\mathrm{Exp}(\kappa)$. When $\tau$ is reached, the process stops and freezes in $\mathbf{p}(\tau)$:
\begin{equation*}
\mathbf{p}^T(t)=
\begin{cases}
\mathbf{p}^T(0)e^{Qt} & \text{if }t\leq \tau\\
\mathbf{p}^T(\tau) & \text{if } \tau<t.
\end{cases}
\end{equation*}
In such a case the final state of the system becomes a random variable so in order to calculate the average final distribution, one has to average $\exp(Q\tau)$ over the distribution $\mathrm{Exp}(\kappa)$. Since $Q$ is bounded, this expectation value exists and is equal to
\begin{equation*}
\kappa \int_{0}^{\infty}e^{Q\tau}e^{-\kappa \tau}d\tau=-\kappa \mathcal{R}(\kappa;Q),
\end{equation*}
where $\mathcal{R}(\kappa;Q)=(Q-\kappa)^{-1}$ is the resolvent of $Q$. Since $Q$ is negative semidefinite, the resolvent is well defined for all strictly positive $\kappa$~\cite{Liggett}. The average probability that a system initially being at state $n$ is found in state $m$ after the stopping  is equal to the $nm$ entry of the resolvent multiplied by $-\kappa$. This shows that $-\kappa\mathcal{R}(\kappa;Q)$ is a stochastic matrix i.e.~the entries in each of its rows are non-negative and sum up to one.

Consider now a Markov process with a non-isolated absorbing state labeled by $0$ with all other states in $\mathcal{S}$ labeled by a sequence of integers increasing from $1$ to $|\mathcal{S}|$. Denote by $\mathcal{\widehat{S}}$ and $\widehat{Q}$ the state space and the infinitesimal generator of the Markov process arising after the elimination of the absorbing state in $\mathcal{S}$. The infinitesimal generator $Q$ has the following form 
\begin{equation*}
Q=\begin{pmatrix}
0 & \mathbf{0}^T \\
\mathbf{a} & A\\
\end{pmatrix},
\end{equation*}
where $\mathbf{0}$ and $\mathbf{a}$ are vectors in $\ell^2(\mathcal{\widehat{S}})$ and $A$ is a bounded linear operator acting on $\ell^2(\mathcal{\widehat{S}})$. Furthermore, $\mathbf{a}$ contains only non-negative entries and
\begin{equation*}
A=\widehat{Q}-\sum_{k=1}^{|\widehat{S}|}a_kP_k,
\end{equation*}
where $P_k$ is the (minimal orthogonal) projection corresponding to the $k$th state of $\widehat{S}$, that is $P_k=\mathbf{e}^{\ }_k\mathbf{e}_k^T$, where $\mathbf{e}^{\ }_k$ is the $k$th canonical unit vector. A short calculation gives
\begin{equation*}
e^{Qt}=\begin{pmatrix}
1 & \mathbf{0}^T \\
\mathbf{b}(t) & e^{At}\\
\end{pmatrix},
\end{equation*}
where
\begin{equation}
\mathbf{b}(t)=\sum_{n=1}^{\infty}\frac{t^{n}}{n!}A^{n-1}\mathbf{a}.
\label{b_def}
\end{equation}
Let $\mathbf{x}\in \ell^2(\widehat{\mathcal{S}})$ be arbitrary and denote the usual scalar product in $\ell^2(\mathcal{\widehat{S})}$ by $(\cdot,\cdot)$ then
\begin{equation*}
(\mathbf{x},A\mathbf{x})=(\mathbf{x},Q\mathbf{x})-\sum_{k=1}^{|\widehat{\mathcal{S}}|}|x|_k^2a^{\ }_k.
\end{equation*} 
Using the facts that $Q$ is negative semidefinite, the absorbing state is not isolated and all the $a_k$'s are non-negative we get $(\mathbf{x},A\mathbf{x})<0$ for all $\mathbf{x}\in \ell^2(\widehat{\mathcal{S}})$, that is $A$ is negative-definite so its inverse exists. This enables us to write \ref{b_def} as
\begin{equation*}
\mathbf{b}(t)=A^{-1}\left (e^{At}-1\right )\mathbf{a}=\left (e^{At}-1\right )A^{-1}\mathbf{a},
\end{equation*}
and the resolvent of $Q$ is
\begin{equation*}
\mathcal{R}(\kappa;Q)=\begin{pmatrix}
\kappa^{-1} & \mathbf{0}^T \\
\mathbf{c}(\kappa) & \mathcal{R}(\kappa;A)\\
\end{pmatrix},
\end{equation*}
where 
\begin{equation}
\mathbf{c}(\kappa)=A^{-1}\left (\mathcal{R}(\kappa;A)-\kappa^{-1}\right )\mathbf{a}.
\end{equation}

Now we restrict our consideration to the lumped SIS dynamics driven on $K_N$ and $S_N$.

\subsubsection{Complete graph}
The Markov process associated with the lumped SIS dynamics on $K_N$ is a birth-death process and, as depicted in Fig.\ref{FIG1}, it has a unique absorbing state. Given the resolvent, if initially there was only one infected individual the average number of infected individuals after the dissolution of the group  is 
\begin{equation}
I(Q)=-\sum_{k=0}^{N}k\kappa \mathcal{R}_{1k}(\kappa;Q)=-\sum_{k=1}^{N}k\kappa \mathcal{R}_{1k}(\kappa;A).
\label{EQ:I1}
\end{equation}
The probability of the epidemic process resulting in more than one infected individuals (which we identify with the probability of an outbreak) is
\begin{equation}
p_{\mathrm{OB}}(Q)=1-(-\kappa\mathcal{R}_{10}(\kappa;Q))=-\sum_{k=1}^{N}\kappa \mathcal{R}_{1k}(\kappa;A),
\label{EQ:P1}
\end{equation}
where we relied the fact that $-\kappa\mathcal{R}_{10}(\kappa;Q)$ is a stochastic matrix. Thus $\mathcal{R}(\kappa;A)$ seems to be the relevant quantity to determine the epidemic threshold of the process. Following \cite{CatorVanMieghem2013}, where the mSIS dynamics was introduced whose stationary distribution can be interpreted as metastable states of the SIS process, we show how the resolvent of $A$ is connected to the $\mathcal{R}(\kappa,\widehat{Q})$, where $\widehat{Q}$ is the infinitesimal generator of the mSIS process associated to $Q$. In order to do so first note that the absorbing state is connected to the rest of the states in $\mathcal{S}$ via a single state. Since this is the $1$ state, i.e.~$\mathbf{a}\propto \mathbf{e}_1$ and the $1\rightarrow 0$ transition rate is $\delta$, the $1$th row of $\mathcal{R}(\kappa;A)$ can be directly computed from $\widehat{Q}$ using the resolvent identity
\begin{equation}
\mathcal{R}(\kappa;\widehat{Q})-\mathcal{R}(\kappa;A)=\mathcal{R}(\kappa;\widehat{Q})(A-\widehat{Q})\mathcal{R}(\kappa;A)
\label{RES}
\end{equation}
that gives
\begin{equation*}
\mathbf{e}^T_1\mathcal{R}(\kappa;\widehat{Q})-\mathbf{e}^T_1\mathcal{R}(\kappa;A)=-\delta\mathbf{e}^T_1\mathcal{R}(\kappa;\widehat{Q})\mathbf{e}^{\ }_1\mathbf{e}_1^T\mathcal{R}(\kappa;A),
\end{equation*}
so
\begin{equation*}
\mathbf{e}^T_1\mathcal{R}(\kappa;A)=\frac{\mathbf{e}^T_1\mathcal{R}(\kappa;\widehat{Q})}{1-\delta\mathcal{R}_{11}(\kappa;\widehat{Q})},
\end{equation*}
which, again using the fact that $-\kappa\mathcal{R}(\kappa;\widehat{Q})$ is a stochastic matrix gives 
\begin{equation*}
p_{\mathrm{OB}}(Q)=\frac{1}{1-\delta\mathcal{R}_{11}(\kappa;\widehat{Q})}.
\end{equation*}
The average number of infected individuals soon after the dissolution is
\begin{equation*}
I(Q)=\frac{-\sum_{k=1}^{N}\kappa k\mathcal{R}_{1k}(\kappa;\widehat{Q})}{1-\delta\mathcal{R}_{11}(\kappa;\widehat{Q})}=\frac{I(\widehat{Q})}{1-\delta\mathcal{R}_{11}(\kappa;\widehat{Q})}.
\end{equation*}

\subsubsection{Star graph}
We are interested in calculating $p_{\mathrm{OB}}(Q)$ and $I(Q)$ in an infection process initiated by one node of the star $S_N$. Whether the infection starts at the perimeter or at the center of the star, a $0$ or a $1$ valued superscript is appended to the previous quantities. As depicted in Fig.~\ref{FIG1}, the absorbing state of the lumped Markov process connects to the rest of the state space by the states $a_0=(0,1)$ and $a_1=(1,0)$ which are the initial states of the infection processes of the aforementioned scenarios. The corresponding rows of $\mathcal{R}(\kappa;A)$ are $\mathbf{e}_{a_0}^T\mathcal{R}(\kappa;A)$ and $\mathbf{e}_{a_1}^T\mathcal{R}(\kappa;A)$. Using the resolvent identity \ref{RES} twice leads to the linear equation
\begin{equation*}
\begin{pmatrix}
\mathbf{e}^T_{a_0}\mathcal{R}(\kappa;\widehat{Q}) \\
\mathbf{e}^T_{a_1}\mathcal{R}(\kappa;\widehat{Q}) \\
\end{pmatrix}
=
L
\begin{pmatrix}
\mathbf{e}^T_{a_0}\mathcal{R}(\kappa;A) \\
\mathbf{e}^T_{a_1}\mathcal{R}(\kappa;A) \\
\end{pmatrix}
\end{equation*}
where
\begin{equation*}
L=
\begin{pmatrix}
1-\delta\mathcal{R}_{a_0a_0}(\kappa;\widehat{Q}) &   -\delta\mathcal{R}_{a_0a_1}(\kappa;\widehat{Q})\\
 -\delta\mathcal{R}_{a_1a_0}(\kappa;\widehat{Q}) &  1-\delta\mathcal{R}_{a_1a_1}(\kappa;\widehat{Q})\\
\end{pmatrix}.
\end{equation*}
After the calculation of $L^{-1}$
\begin{equation*}
L^{-1}=
\frac{1}{\mathrm{det}(L)}
\begin{pmatrix}
1-\delta\mathcal{R}_{a_1a_1}(\kappa;\widehat{Q}) &   \delta\mathcal{R}_{a_0a_1}(\kappa;\widehat{Q})\\
 \delta\mathcal{R}_{a_1a_0}(\kappa;\widehat{Q}) &  1-\delta\mathcal{R}_{a_0a_0}(\kappa;\widehat{Q})\\
\end{pmatrix},
\end{equation*}
and using the stochasticity of $-\kappa\mathcal{R}(\kappa,\widehat{Q})$ we arrive at
\begin{equation}
\begin{pmatrix}
p^{(0)}_{\mathrm{OB}}(Q) \\
p^{(1)}_{\mathrm{OB}}(Q) 
\end{pmatrix}
=
L^{-1}
\begin{pmatrix}
1\\
1\\
\end{pmatrix}.
\label{EQ:P2}
\end{equation}
The corresponding expression regarding $I^{(0)}(Q)$ and $I^{(1)}(Q)$ is 
\begin{equation}
\begin{pmatrix}
I^{(0)}(Q) \\
I^{(1)}(Q) 
\end{pmatrix}
=
L^{-1}
\begin{pmatrix}
I^{(0)}(\widehat{Q}) \\
I^{(1)}(\widehat{Q})\\
\end{pmatrix},
\label{EQ:I2}
\end{equation}
where the expectation values $I^{(0)}(\widehat{Q})$ and $I^{(1)}(\widehat{Q})$ of the mSIS process can be calculated by 
\begin{align*}
I^{(0)}(\widehat{Q})&=-\kappa\sum_{k=1}^{N-1}\sum_{l=0}^1(k+\delta_{l,1})\mathcal{R}_{(0,1),(l,k)}(\kappa,\widehat{Q}),\\
I^{(1)}(\widehat{Q})&=-\kappa\sum_{k=1}^{N-1}\sum_{l=0}^1(k+\delta_{l,1})\mathcal{R}_{(1,0),(l,k)}(\kappa,\widehat{Q}).
\end{align*}
The previous expressions can be used directly to calculate $I^{(0)}(Q)$ and $I^{(1)}(Q)$ which are the main objects of investigations in the next Section. 
\section{Discussion}
\label{DISCUSSION}
Consider a finite group of $N$ individuals. Since the infinitesimal generator $Q$ of the SIS dynamics has a vanishing row sum, zero is an eigenvalue of $Q$ (the corresponding right eigenvector is proportional to $(1,\dots,1)^T$). As a consequence $\mathcal{R}(\kappa;Q)$ has a pole in $\kappa=0$ and its entries diverge to $-\infty$ when zero is approached from the right of the real axis. This fact guarantees that, by the vanishing right hand sides of Eq.\ref{EQ:I1}, \ref{EQ:P1}, \ref{EQ:P2} and \ref{EQ:I2}, in the $\kappa\rightarrow 0$ limit we can get back the long time behaviour of the usual SIS model.

From a practical point of view the interesting regime of the parameter space formed by $\beta$, $\delta$ and $\kappa$ is characterized by $\kappa\ll\beta,\delta$. This is the case when the time scale of the spreading process is much smaller than the time scale of the decay of the group in which the spreading occurs. We note that - as it is usual - one of the three parameters can be always eliminated by using an independent combination of the ratios of the rates instead of using the rates themselves.  We, however, omit this option and keep all the parameters. From now on, a section of a function of several parameters obtained by fixing all its dependancies except one will be denoted by a subscript of the running parameter. 

\begin{figure}[!ht]
\centering
\includegraphics[scale=0.70,trim={0 0 0 0},clip]{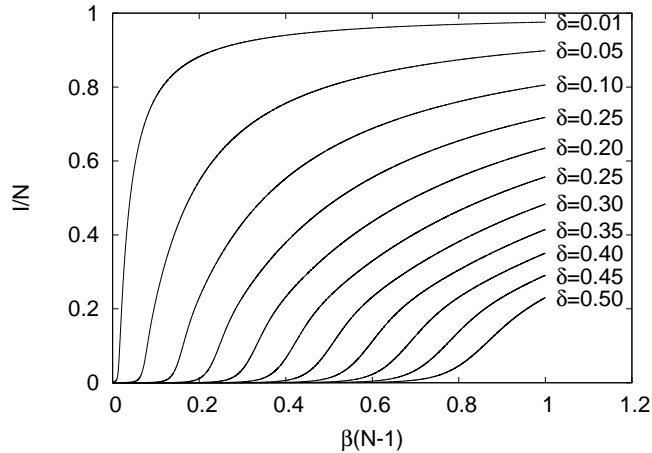}
\caption{Average final density of infected individuals in an epidemic process on $K_{50}$ initiated by one infected individual. The resolution of the numerical calculations is $\Delta \rho=10^{-4}$ and $\kappa=10^{-3}$.}
\label{FIG3}
\end{figure}

In order to study the effect of the finite lifetime on the final state of the SIS process on $K_N$, we set $\kappa=10^{-3}$ and performed numerical calculation of the resolvent to obtain $I_{\beta}(Q)/N$ for various values of $\delta$. The results can be seen in Fig.~\ref{FIG3}, where we plot  the density of infected individuals against the strength of the infection. The strength of the infection $\rho$ is the rescaled form of $\beta$ that enables us to compare SIS dynamics taking place in populations that have different sizes. Since the initial state of the process contains only one individual it seems plausible to choose $\rho=\beta(N-1)$. It is surprising - at least to us - that even a small decay rate can imply a drastic change in the long time behaviour. 
Assuming a fixed $\delta$, the small value of $\rho$ leads to small $I_{\rho}(Q)/N$. At the beginning, increasing $\rho$ has only a little effect on $I_{\rho}(Q)/N$ then close to $\delta$ the situation changes and $I_{\rho}(Q)/N$ starts to grow very fast. This rapid growth has two different phases: an initial convex phase pushes out the density from the close-to-zero value which after the forecoming concave phase starts and leads to a moderate grow in $I_{\rho}(Q)/N$.

\begin{figure}[!ht]
\centering
\includegraphics[scale=0.70,trim={0 0 0 0},clip]{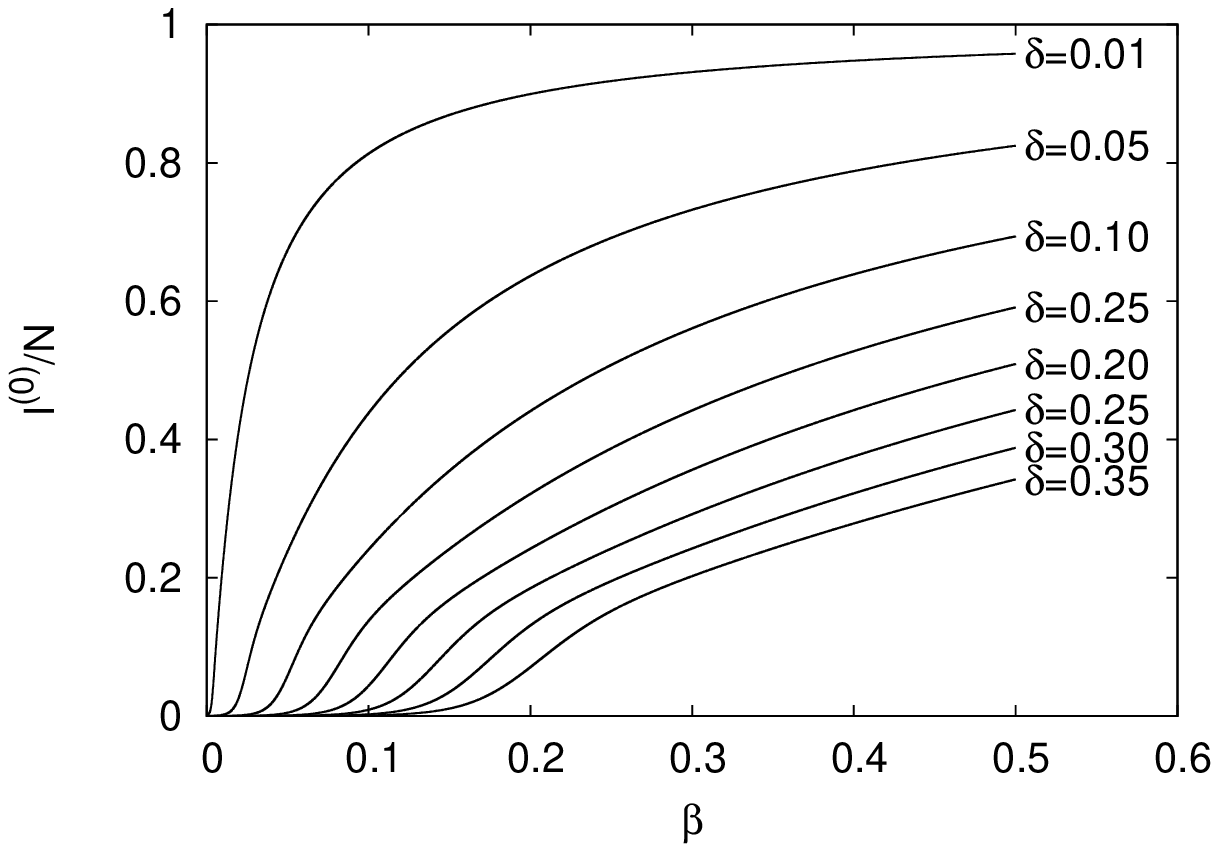}
\caption{Average final density of infected individuals in an epidemic process on $S_{50}$ initiated by one infected leave. The resolution of the numerical calculations is $\Delta \rho=10^{-4}$ and $\kappa=10^{-3}$.}
\label{FIG4}
\end{figure}

\begin{figure}[!ht]
\centering
\includegraphics[scale=0.70,trim={0 0 0 0},clip]{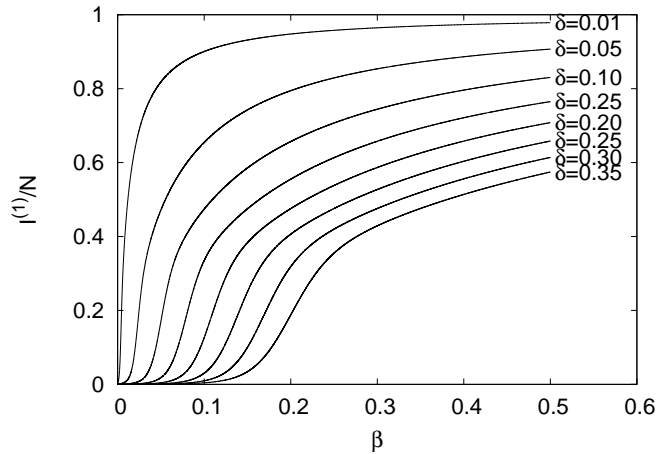}
\caption{Average final density of infected individuals in an epidemic process on $S_{50}$ initiated by the infected center node. The resolution of the numerical calculations is $\Delta \rho=10^{-4}$ and $\kappa=10^{-3}$.}
\label{FIG5}
\end{figure}

In the case of $S_N$, the calculation of $I_{\rho}^{(0)}(Q)$ and $I_{\rho}^{(1)}(Q)$ with the choice $\rho=\beta$ leads to roughly the same experience (see Fig.~\ref{FIG4} and Fig.~\ref{FIG5}) but the region of $\rho$ where $I_{\rho}(Q)/N$ starts to grow fast is far away from the neighborhood of $\delta$. The reason behind this is the fact that the natural scaling of $\rho/\beta$ in the case of $S_N$ is different from the unit. In the case of the mSIS dynamics it can be shown \cite{CatorVanMieghem2012} that the scaling is of the form $\rho/\beta\propto \alpha(N)/\sqrt{N-1}$, where the term $\alpha(N)$ is of order $\ln^{1/2}(N-1)$. Unfortunately such a scaling seems to be inappropriate in explaining these curves and we have not yet been able to find the appropriate scaling law.
The main difference between spreading processes initiated by the center and one of the leaves of the star is the intensity of the process. This means that at a given fixed value of $\beta$ and $\delta$ the density of infected individuals in the former case is about twice as great as it would be in the latter case. This is understandable by the structure of the star graph - the infection spreads very fast through the central node once it has become infected.
\begin{figure}[!ht]
\centering
\includegraphics[scale=0.75,trim={1cm 0 0 0},clip]{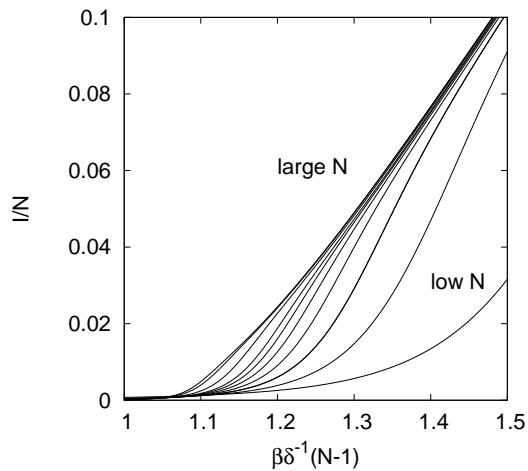}
\caption{The large $N$ dependence of $I(Q)/N$ on the graph $K_N$. The curing rate is fixed to $\delta=2.5\cdot10^{-2}$ and $\kappa=10^{-3}$.}
\label{FIG6}
\end{figure}
In order to gain insight in the $N\rightarrow \infty$ limit we fixed $\kappa=10^{-3}$ and $\delta=2.5\cdot 10^{-2}$ and evaluated the $\rho$ dependence of the density of infected individuals for various increasing values of population size. In Fig.~\ref{FIG6} the case of $K_N$ is illustrated and in Fig.~\ref{FIG7} the results for the case of $S_N$  are shown. It can be seen that in both cases the width and the height of the convex phase of $I_{\rho}(Q)$ becomes smaller and smaller as $N$ increases. On the other hand the concave phases seemingly accumulate in the large $N$ limit. For sufficiently large values of $\rho$, that are far away from the points where these curves separate from the horizontal axis, the difference between the curves corresponding to different population sizes becomes less and less pronounced. Similar phenomenon appears in the case of $S_N$ as well. It is not clear whether the left directed shift of the place of the separation of $I_{\rho}(Q)$ and $I^{(0)}_{\rho}(Q)$ from the horizontal axis saturates  with increasing $N$ or not. The absence of this kind of behaviour would be a sign of a well-defined critical infection strength in the large $N$ limit.
\begin{figure}[!ht]
\centering
\includegraphics[scale=0.75,trim={1cm 0 0 0},clip]{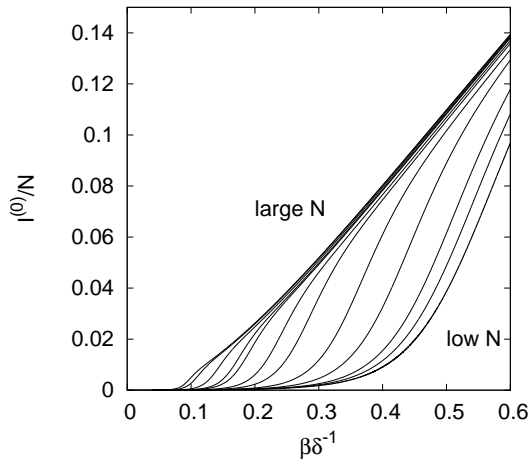}
\caption{The large $N$ dependence of $I^{(0)}(Q)/N$ on the graph $S_N$. The curing rate is fixed to $\delta=2.5\cdot10^{-2}$ and $\kappa=10^{-3}$.}
\label{FIG7}
\end{figure}

\begin{figure}[!ht]
\centering
\includegraphics[scale=0.70,trim={0 0 0 0},clip]{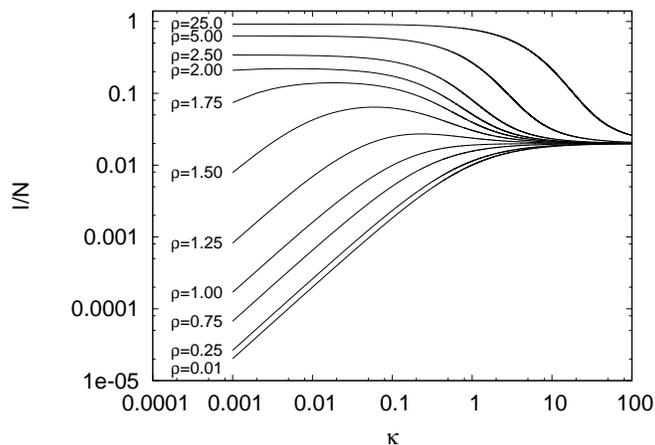}
\caption{The large $\kappa$ dependence of $I(Q)/N$ on the graph $K_{50}$. The curing rate is fixed to $\delta=1$ and $\Delta\kappa=10^{-3}$.}
\label{FIG8}
\end{figure}

Finally we investigated the robustness of our results by examining the $I_{\kappa}(Q)/N$ curves (curing and infection rates are fixed) of $K_N$. Results are depicted in Fig.~\ref{FIG8}. 
As $\kappa$ tends to infinity, the probability that any jump in states space occurs between the starting time of the dynamics and the dissolution of the group tends to zero. This means that as $\kappa\rightarrow\infty$, we have $I_{\kappa}(Q)/N\rightarrow 1/N$. This can be clearly seen at the large $\kappa$ region of the diagram which shows interesting structure only in the $\kappa\lesssim \delta,\beta$ region. There, the curves clearly present the aforementioned behaviour. For small values of $\rho$, an increase in the infection strength results in a moderate grow of $I_{\rho}(Q)/N$, but at a suitable value the convex phase of growing starts. This part of the curves seems to be sensitive to the value of $\kappa$. Surprisingly, the concave region of $I_{\rho}(Q)/N$ seems to be very roboust. 
\begin{figure}[!ht]
\centering
\includegraphics[scale=0.70,trim={0 0 0 0},clip]{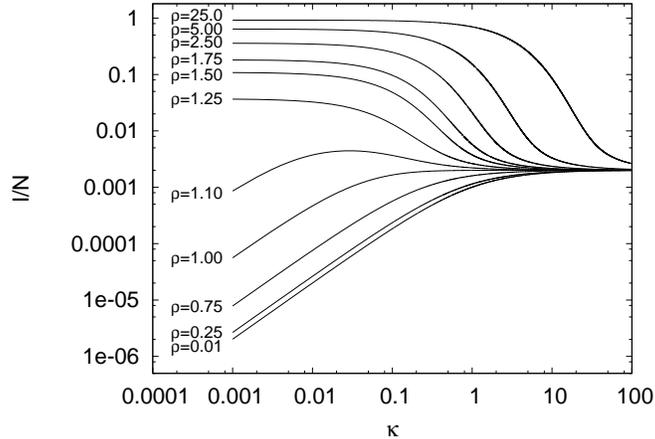}
\caption{The large $\kappa$ dependence of $I(Q)/N$ on the graph $K_{500}$. The curing rate is fixed to $\delta=1$ and $\Delta\kappa=10^{-3}$.}
\label{FIG9}
\end{figure}
Since a previous observation suggested that the convex part of the $I_{\rho}(Q)/N$ curve seems to disappear in the $N\rightarrow\infty$ limit we repeated the whole calculation of the $I_{\kappa}(Q)/N$ curves with $N=500$. The results are depicted in Fig.~\ref{FIG9}. It can be clearly seen that a larger $N$ value supports a larger interval of $\rho$ where $I_{\rho}(Q)/N$ has only a weak dependence on (the sufficiently small) $\kappa$.

\section{Conclusion}
\label{CONCLUSION}

In this paper an epidemiological study is presented in a finite size population. The system is modeled by a continuous time finite size Markov chain
without eliminating the absorbing state. The system dynamics is allowed to develop for a finite exponentially distributed duration period. Formulae for the probability of the outbrake are derived and the behaviour of the average final density as a function of the parameters of the SIS
dynamics is investigated. Numerical calculations of these formulae are carried out and analyzed for two different graph types, whose state spaces are lumpable: the complete graph and the star. In the long time limit a threshold like behaviour of the model system is identified.

As next steps we are extending this system to account for concurrent group formation processes parallel with the dynamics of the epidemics.

\section{Acknowledgement}
We thank to Ericsson Ltd. for support via the ELTE CNL collaboration.

\bibliography{finite_lifetime}{}
\bibliographystyle{plain}
\end{document}